\documentclass[10pt]{elsart}
\usepackage{graphicx}
\usepackage{amsmath}
\usepackage{subfigure}
\usepackage{amsfonts}
\usepackage{amssymb}

\newcommand {\be} {\begin{equation}}
\newcommand {\ee} {\end{equation}}
\newcommand {\Be}{\begin{eqnarray*}}
\newcommand {\Ee} {\end{eqnarray*}}
\newcommand {\bey} {\begin{eqnarray}}
\newcommand {\eey} {\end{eqnarray}}

\begin{document}
\begin{frontmatter}

\title{Stretched--exponential relaxation in arrays of coupled rotators}

\author[label1]{Maria Eleftheriou}
\author[label1]{Stefano Lepri}
\author[label3,label1]{Roberto Livi}
\author[label5]{Francesco Piazza}
\ead{Francesco.Piazza@epfl.ch}

\address[label1]{Istituto Nazionale per la  Fisica della Materia, UdR
Firenze, via G. Sansone 1 I-50019 Sesto Fiorentino, Italy}
\address[label3]{Dipartimento di Fisica, via G. Sansone 1 I-50019, Sesto
Fiorentino, Italy}
\address[label5]{Laboratoire de Biophysique Statistique, \'Ecole
Politechnique F\'ed\'erale de Lausanne, SB--ITP, BSP 720, CH--1015, Lausanne,
Switzerland}

\date{\today}

\begin{abstract}

We consider the non--equilibrium dynamics of a chain of classical rotators coupled 
at its edges to an external reservoir at zero temperature.  We find that the energy 
is released in a strongly discontinuous fashion, with sudden jumps alternated with
long stretches during which dissipation is extremely weak. The jumps mark the
disappearance of strongly localized structures, akin to the rotobreather solutions of
the Hamiltonian model, which act as insulating boundaries of a hot central core. As a
result of this complex  kinetics, the ensemble--averaged energy  follows a stretched
exponential law until a residual pseudo--stationary state is attained, where the hot
core has reduced to a single localized object.

We give a statistical description of the relaxation pathway and connect it to
the properties of return periods of rare events in correlated time series.
This approach sheds some light into the microscopic mechanism underlying the
slow dynamics of the system. 

Finally, we show that the stretched exponential law
remains unaltered in the presence of isotopic disorder.

\end{abstract}


\end{frontmatter}

\maketitle

\section{Introduction}

Many physical systems display unusual relaxation properties, such as  slow
kinetics in glasses~\cite{revglass} or  dynamics of
biomolecules~\cite{exampbio}, which are usually associated with the presence of
a complex energy landscape. Long--living transient states also arise in
non--linear  systems, where the relaxation properties may be affected by the
emergence of  localized vibrations like solitons or kinks (see
e.g.~\cite{ruffo}). Of special interest is the role of a recently discovered
class of spatially localized,  time--periodic excitations, termed discrete
breathers~\cite{breath1}.  At variance with their counterparts in the
continuum, they exist  under very general conditions and their stability
properties have been thoroughly investigated~\cite{breath1}.  In particular,
these objects may easily self--excite under very different  physical
conditions, as exemplified by several   numerical and experimental 
studies~\cite{breath1,Ustinov1,Ustinov2,Sievers1,Sievers2,page1,page2}.

A typical situation where breathers form spontaneously can be obtained by
cooling the lattice boundaries, thus driving energy out of the system. In such
conditions one is faced with a genuine nonlinear effect: the  system prevents
the complete dissipation of the energy by storing part of it in a residual
long--living breather or multi--breather state. Such a phenomenon was first
observed in a model of coupled harmonic oscillators pinned by an on--site
nonlinear force~\cite{Aubry1,Aubry2} and later recovered in models of
nonlinearly coupled oscillators without on-site
interaction~\cite{noi,chaos,Reigada}. Nonetheless, the process of energy
dissipation was found to exhibit quite different behaviors in these two classes
of models. The main mechanism yielding a power-law decay of the energy to the
residual state has been described in previous papers by the
authors~\cite{noi,chaos}. The stretched exponential behaviour, primarily
observed in models with on--site interaction, was then conjectured to represent
a typical feature for this class of models. Nonetheless, a satisfactory
understanding of this peculiar decay process was still lacking.

The present paper deals with the nonequilibrium transient dynamics of a model
of coupled classical rotators on a one--dimensional lattice. Two  different
cases will be considered, the one with pure nearest-neighbours coupling and the
one with an additional on--site force which is a generalized  version of the
well-known discrete sine-Gordon chain. The quantum version of the latter model
has been introduced by Fillaux and collaborators~\cite{FC} to describe
inelastic--neutron--scattering experiments on methyl-pyridine crystals. As this
system displays evidence of extremely slow kinetics~\cite{FC2}, it is
particully interesting to investigate  how cooling and nonlinear effects may
affect thermalization. Although limited to the classical case, our study  may
provide some insight on the observed phenomenology.

The second motivation comes from the problem of stationary heat transport in
this class of models. It has been recently recognized that low-dimensional
lattices show anomalous properties, namely that transport coefficients (e.g.
the thermal conductivity) diverge~\cite{physrep}. The analysis of several
models clarified that this striking feature occurs generically whenever
momentum is conserved. For lattice models, this amounts to requiring that at
least one acoustic phonon branch is present in the harmonic limit. Remarkably,
the only known exception is actually the pure nearest--neighbor model, which
displays normal transport~\cite{therm}. Since the transient dynamics is
connected to the system's response, it is interesting to investigate the energy
relaxation and to compare the peculiarities of each class of models. As
we shall see, there are indeed some qualitative differences whose genuine
nonlinear origin (excitation  of long--lived localized structures) points at
significantly differerent transport mechanism than those based on the customary 
phonon theory.

The plan of the paper is as follows. The details of the system of coupled
classical  rotators cooled at its boundaries are presented in Section 2: we
consider both the cases with and without  on-site interaction. The simulations
reported in Section 3 indicate that the energy relaxation dynamics displays
in both cases a  stretched--exponential behavior. The statistical
interpretation presented in Section 4 relates such a stretched--exponential
behavior with rare-event statistics. To our knowledge, this correspondence has
never been pointed out before in the literature. In section 5 we show that the
main features and the interpretation do not change when isotopic disorder
is introduced.

Two relevant results of our analysis merit to be stressed from the very
beginning. The first one concerns the robustness of the phenomenon with respect
to different variants of the model. The second one amounts to put such a
phenomenon in a different perspective: the  stretched--exponential energy decay
process seems to be common to one--dimensional systems whenever the typical
solutions are static breathers, rather than being  a
peculiarity of models with on--site interaction.

\section{The model\label{sec:1}}

We study the relaxation toward equilibrium of a chain of classical rotators
coupled on a lattice with damping acting at its edges. 
The equations of motion are
\begin{equation}
I_i\ddot \phi_{i}= -G\sin\phi_{i}
+ K\left[\sin(\phi_{i+1}-\phi_{i}) +\sin(\phi_{i-1}-\phi_{i})\right]
-\gamma  \dot\phi_i \left[ \delta_{i,1} +\delta_{i,N} \right]
\label{onsite}
\end{equation}
where $\phi_{i}$ in the angle variable of the $i$-th rotator  (with $i=1,2,\ldots
N$). The Hamiltonian version of~(\ref{onsite}) is sometimes referred
to as the sine--lattice equations~\cite{TP,TP2}.
We impose free--end boundary conditions  ($\phi_1=\phi_{0}$,
$\phi_{N}=\phi_{N+1}$)  since we know from previous studies that this choice
favors spontaneous localization~\cite{noi}.
In the following we will also consider the special case $G=0$, where only 
interactions between nearest neighbors are present and refer to it 
as the NN model.  Furthermore, we will set $I_i=1$ for all $i$ and 
$K=1.4911$, a value that in our units corresponds to 
experimental data on 4--methyl--pyridine~\cite{FC}.

The general layout of a simulation has been detailed
previously~\cite{noi,chaos}. First, an equilibrium microstate is generated by
letting the Hamiltonian (microcanonical, i.e. $\gamma$ = 0) system  evolve for
a sufficiently long transient (typically 500 time units).   We used the 5$^{\rm
th}$--order  symplectic Runge--Kutta--Nystr\"{o}m  algorithm~\cite{Calvo}.  The
initial condition for the transient is assigned by setting all angles $\phi_i$
to zero and by drawing velocities at random from a Gaussian distribution with
standard deviation equal to the square root of twice the energy density
$E(0)/N$.  The resulting set of $\phi_i$ and $\dot \phi_i$ is then used as
initial  condition to integrate  Eq.~(\ref{onsite})  with $\gamma > 0$ with a
standard 4$^{\rm th}$--order Runge--Kutta algorithm. The subsequent dissipative
dynamics is described in the following section.

\section{The relaxation dynamics\label{sec:2}}

\begin{figure}[!t]
\centering
\subfigure[]{
\includegraphics[width=5.8 truecm,clip]{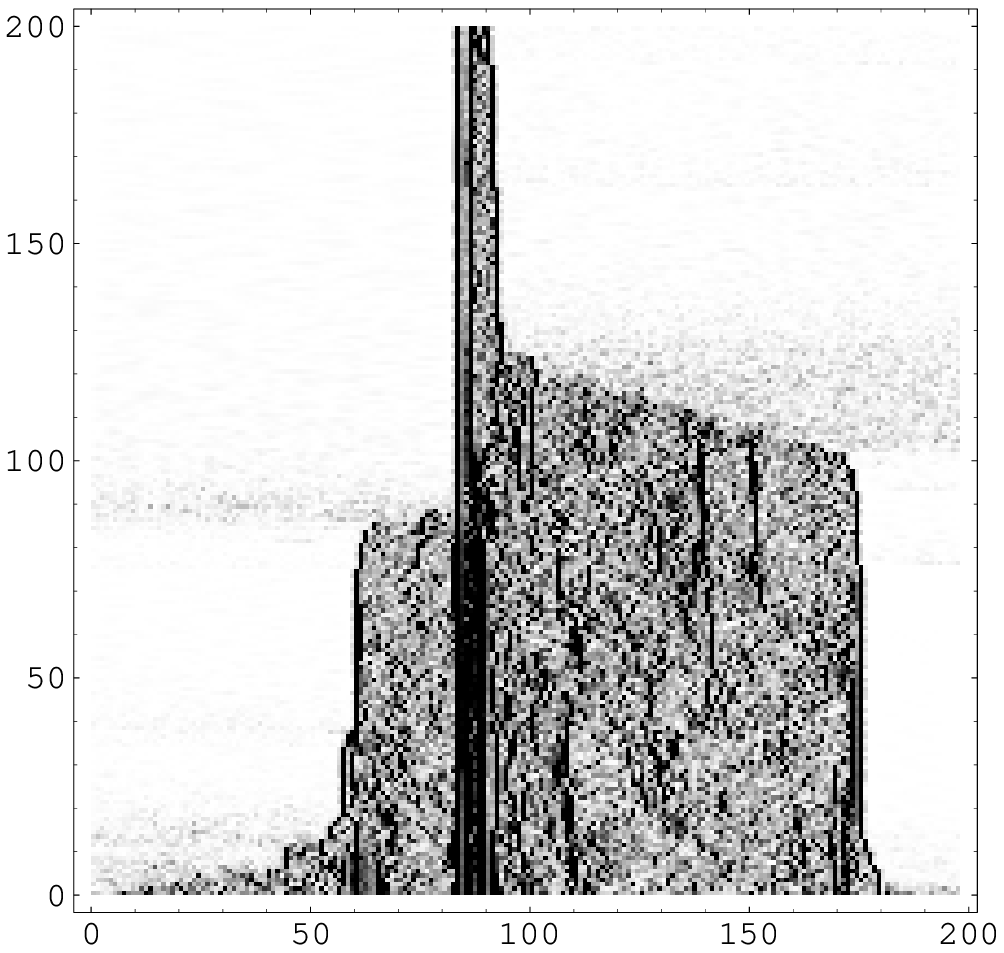}}
\subfigure[]{
\includegraphics[width=7.5 truecm,clip]{Figura2.eps}}
\subfigure[]{
\includegraphics[width=5.8 truecm,clip]{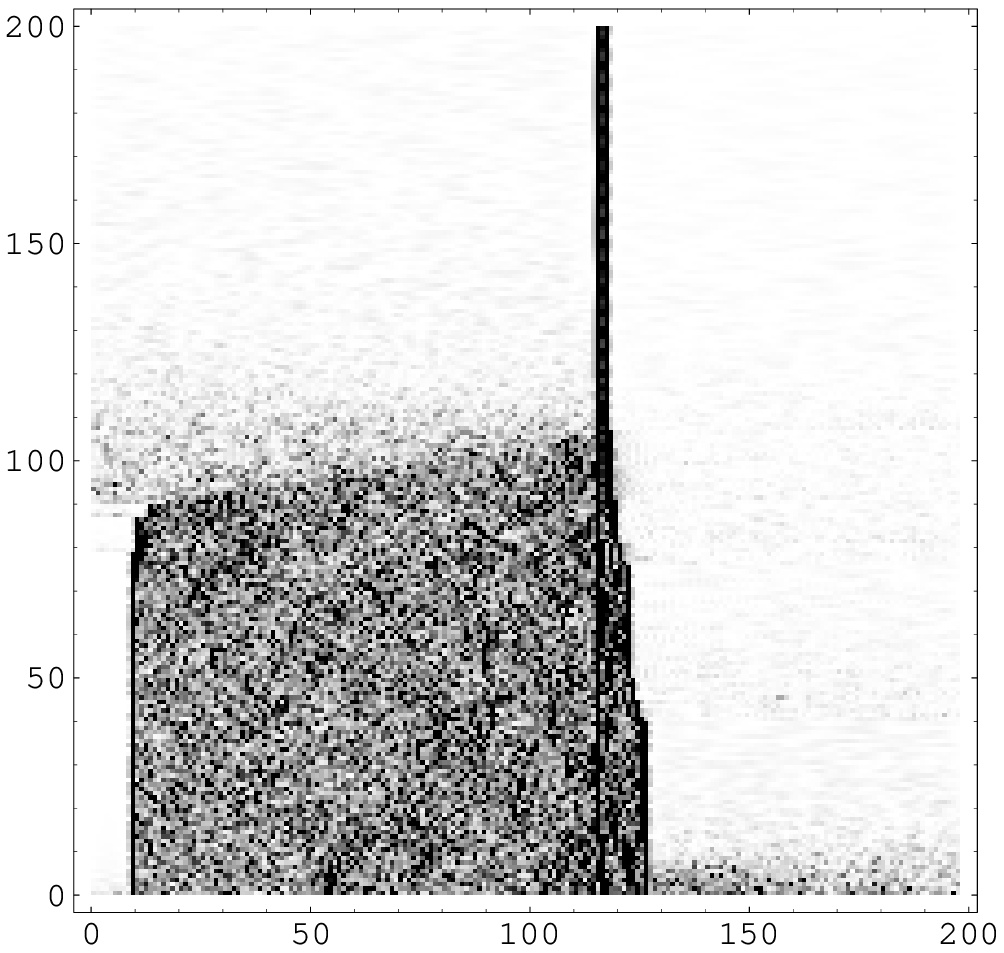}}
\subfigure[]{
\includegraphics[width=7.5 truecm,clip]{Figura4.eps}}
\caption{
Relaxation in a chain with $N=200$, $\gamma=0.1$, $K=1.4911$, $E(0)/N=5$.
Upper panels $G=0$,  lower panels $G=1$.
(a), (c) Space--time  density plots of the symmetrized site energies.
The regions with higher energies are labeled with darker shades
of grey.  
The chain sites $i$ are reported on the horizontal axis ($2 \leq i \leq N-1$), 
the vertical axis describes the time evolution in units of sampling 
times (250 natural time units).
(b), (d) Decay of the normalized lattice energies. The time
is expressed in units of the sampling time of the lattice site energies.}
\label{fig10}
\end{figure}

A typical simulation is illustrated in Fig.~\ref{fig10}.   In the left panels we show  
the space--time contour plots of the symmetrized site energies $e_i$, defined as
\begin{equation}
\label{hsite}
e_i \; = \;\frac{1}{2} \dot{\phi}_i^2 +
         \frac{1}{2}
       K\left[ 2-\cos(\phi_{i+1}-\phi_i) -\cos(\phi_{i}-\phi_{i-1} ) \right]
       + G[1-\cos\phi_i] \quad .
\end{equation}
In the right panels we report the  time series of the total normalized
energy $E(t)/E(0)$ for the same simulation runs.

The energy density contour plot shows that the process is associated with 
the spontaneous creation of long--lived localized objects, characterized by 
a fast librating motion of a single rotator almost decoupled from its neighbors.
These localized excitations  (rotobreathers) have already been shown to exist in
the Hamiltonian lattice~\cite{TP}.  
The appearance of such
strongly localized objects is due to the choice of the coupling
which is of the same order as the on--site term (see the analysis
in Ref.~\cite{TP}). 

The two outermost rotobreathers systematically act as barriers between
a central ``hot'' region of the chain and its boundaries, thus blocking
the energy flow toward the environment. At some stage, such dynamical barriers
are spontaneously destabilized, thus allowing a portion of the trapped 
energy to rapidly flow away. 
This process goes on progressively reducing the size of the central hot patch
until a single rotobreather survives in the bulk.    
We checked that its motion  is periodic by computing the spatiotemporal 
spectrum $S(k,\omega)$ of the $\phi_i$s   for several wave-numbers $k$.
Indeed $S(k,\omega)$ displays sharp peaks at the fundamental frequencies plus
small--amplitude harmonics.

The right panels of Fig.~\ref{fig10} illustrate how the energy flows toward the 
reservoirs. In accordance with the above described phenomenology, the energy decay
proceeds in a characteristic step--wise fashion, through a series of sudden jumps 
separated by long, approximately constant plateaus. 
The jumps correspond to the disappearance of one of the rotobreathers at the
boundaries of the hot region. Between two consecutive jumps, the energy remains
approximately constant with  very small fluctuations. 

\begin{figure}[!t]
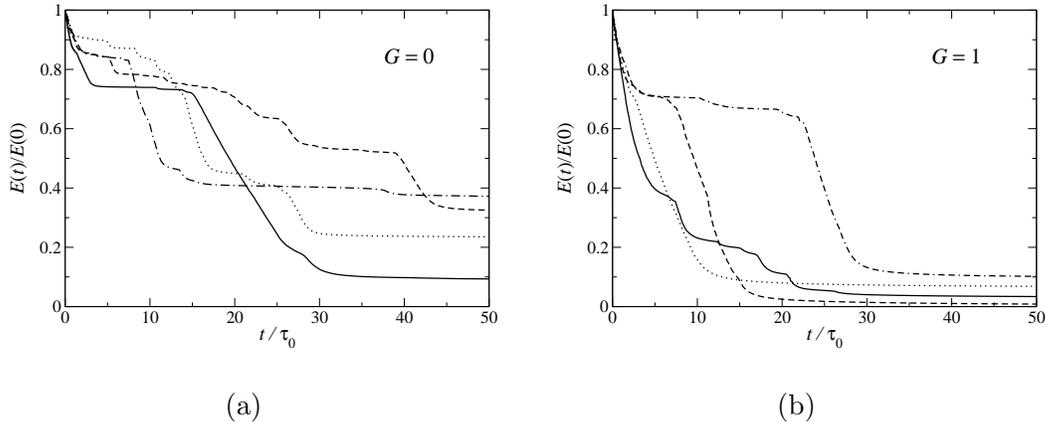

\centering
\subfigure[]{
\includegraphics[width=6.5cm,clip]{Figura5.eps}}
\hspace{.5 cm}
\subfigure[]
{\includegraphics[width=6.5cm,clip]{Figura6.eps}}
\caption{Decay of the normalized total energy for four different 
initial conditions with $E(0)/N = 5$. (a) NN model. (b) Full
sine--lattice model.
The lattice size is $N=200$. Other parameters are $\gamma=0.1$, 
$K=1.4911$.}
\label{f:enall}
\end{figure}

Our simulations show that this kind of scenario persists also
in the case $G=0$, thus confirming previous qualitative results~\cite{russi}. 
This finding is also illustrated in Fig.~\ref{f:enall}, where we compare 
the energy decay of four different initial conditions for the two 
kinds of potentials.  In order to emphasize the slow character of 
the process, we rescale the time by  $\tau_0=N/2\gamma$, which has
been shown to be the characteristic time scale over which 
edge dissipation occurs~\cite{noi}.

It is interesting to observe that different initial 
conditions may give rise  to substantially different sequences
of jumps and plateaus, that may be regarded as different realizations
of a stochastic process. For comparison, we remark that such sensitivity to the choice 
of the initial condition has not been observed in the relaxation of other 
non--linear systems, where different stories approximately resemble
each other and display a much smoother decay~\cite{noi,chaos}.

\begin{figure}[!t]
\centering
\subfigure[]{
\includegraphics[width=10. cm,clip]{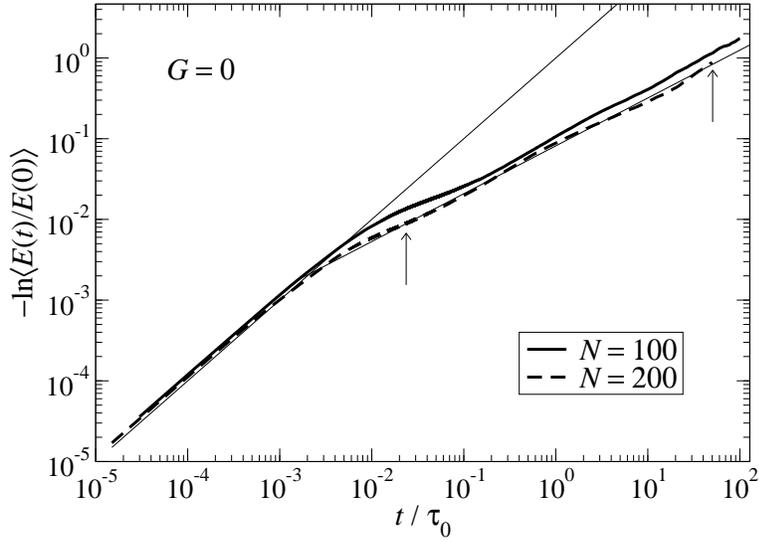}}
\hspace{.5 cm}
\subfigure[]
{\includegraphics[width=10. cm,clip]{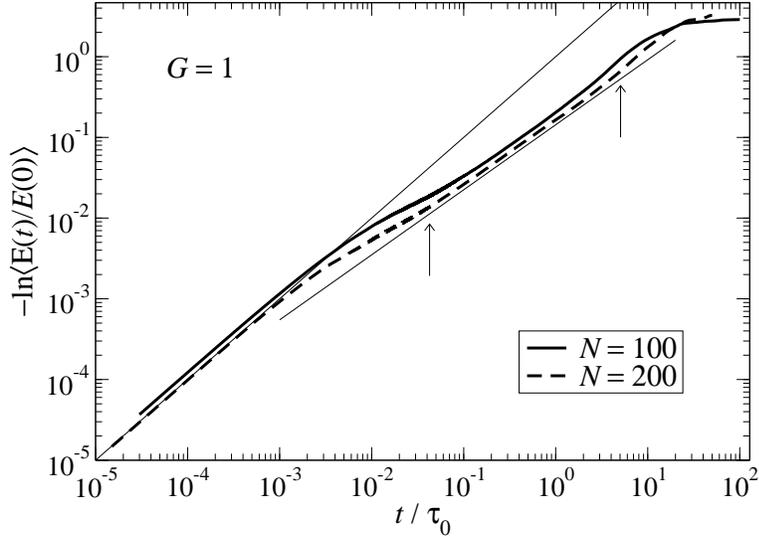}}
\caption{Decay of the normalized total energy averaged over a set
of  99 initial conditions.
We plot $-\ln \langle E(t)/E(0) \rangle$ in log--log scale
with $E(0)/N = 5$ and for two different 
lengths of the chain. 
The  thin lines correspond to $\exp[-t/\tau_0]$ (leftmost curve),
and to a stretched exponential fit $\exp[-(t/\tau)^\sigma]$
of the data between the two arrows.
(a) NN model, $\sigma = 0.59$. (b) Full
sine--lattice model $\sigma = 0.80$.
Other parameters are $\gamma=0.1$,  $K=1.4911$.}
\label{f:stretch}
\end{figure}

In order to give  a more quantitative description of the relaxation process,
it is very interesting to study the average behavior of the energy decay.
Figure~\ref{f:stretch} shows that the latter is characterized by three different
scaling regimes. These are best visualized by plotting the quantity
$-\ln \langle E(t)/E(0) \rangle$ in log--log scale, where a law
of the form $\exp(-t^\sigma)$ shows up as a straight line with slope $\sigma$.
We find that both the systems with $G=0$ and $G =1$ exhibit the same 
properties.
At short times ($t \ll \tau_0$), the energy decays exponentially as 
$E(t)=E(0)\exp(-t/\tau_0)$.
This is what one observes for a harmonic lattice of the same size,
thus signaling that the non--linear effects are negligible on such a time scale.
Actually, the same holds for the Fermi--Pasta--Ulam chain~\cite{noi}.
At intermediate times ($t \simeq 10^{-2}\tau_0$), 
the curves undergo a change of slope and slow 
down to stretched exponential laws. The  knee associated with such crossover
may  be due to power--law corrections, which are known to be present at this stage
in other systems~\cite{noi}.
Eventually ($t > 10 \tau_0$), the residual quasi--stationary state is reached.
In this last stage,  a finite fraction of the initial energy is stored 
in a single rotobreather and remains constant over the rest of the simulation.

In Figure~\ref{f:stretch} we also consider the effects of the lattice size 
for the same value of the damping constant $\gamma$. In order to make the
comparison, we rescale the time axis by the size--dependent constant 
$\tau_0=N/2\gamma$. We see that the data for $N=100,200$ collapse fairly 
well for $t < \tau_0$ and follow approximately the same trend at later times.
More marked differences are observed in the first crossover region.
In particular, the range of stretched exponential decay seems to be larger
for the longer chains.

\begin{table}
\caption{Dependence of the exponent $\sigma$ on the initial energy density 
for $N=100$ and $\gamma=0.1$.\label{t:exp}}
\begin{center}
\begin{tabular}{ccccc}
\hline
$E(0)/N$ & \ \ \ \ \ & $G=0$ & \ \ \ \ \ & $G=1$ \\
\hline\hline
 3   & \ \ \ \ \   & 0.88  & \ \ \ \ \ & 0.93  \\
 5   & \ \ \ \ \   & 0.59  & \ \ \ \ \ & 0.80  \\
 7   & \ \ \ \ \   & 0.41  & \ \ \ \ \ & 0.63  \\
 8   & \ \ \ \ \   & 0.40  & \ \ \ \ \ & 0.49  \\
\hline
\end{tabular}
\end{center}
\end{table}

To complete this analysis, we checked the dependence of the exponent $\sigma$  on the
initial energy. The data reported in Table~\ref{t:exp} show that   $\sigma$ decreases
by increasing  $E(0)/N$. Accordingly, this suggests that the higher the initial energy
density, the longer is the ``lifetime'' of the rotobreathers that appear in the
system.  There are some results in the literature that support this inference for the
sine--lattice systems (with and without local coupling).  In fact,  it has been  found
that the lifetime of an excitation of energy $E$ in a nonlinear system  should be
proportional to $\exp\sqrt{E}$~\cite{Kaneko}.

Finally, we wish to mention that a similar stepwise decay process yielding 
an average stretched--exponential law has
been found also in two--dimensional easy--plane ferromagnets~\cite{Bishop}.
In such case, annihilation of vortex--antivortex pairs is the basic physical
mechanism.

\section{Statistical analysis\label{sec:3}}

In this section we attempt to understand better the origin of the
slow kinetics. In particular, we shall focus on the statistics of
plateaus and energy jumps during energy relaxation. As we already
noticed above, the total energy of the lattice  decreases in a
stepwise fashion. Between subsequent jumps no appreciable energy
exchange with the external world occurs and a plateau is observed
(see again Fig.~\ref{f:enall}), whose duration may vary over different time 
scales. For a  single realization of
a given initial condition,  we can thus
approximate the curve $E(t)$ as a piece-wise function 
through the sequences $\Delta E_n$ (energy
jump at the $n$-th step) and $\Delta t_n$ (duration of the $n$-th
plateau):
\begin{equation}
\label{piecewdec}
\frac{E(t)}{E(0)} = 1-\sum_{n=1}^{N_p}
                                       \Delta E_n
                                       \Theta  \left(
                                       t - \sum_{m=0}^{n-1}
                                       \Delta t_m
                                       \right) \quad ,
\end{equation}
where $\Theta(x)$ is the Heaviside step function.
In the spirit of a statistical description we thus
regard each sequence $(\Delta E_n, \Delta t_n)$ for $n=1, \ldots
N_p$ as a realization of a random process.

Since all the features of this process are unknown, we devised a
numerical procedure to extract from the simulations both the
$\Delta t_n$ and $\Delta E_n$ for each realization.
The first task is to locate the beginning and end of each plateau to a
reasonable level of accuracy.  We chose to sample the whole
time-series (with an adjustable sampling interval $t_s$) and look
for the times where the energy differences $J(t) = E(t+t_s)-E(t)$
exceed some pre--assigned threshold value $|J(t)|>q$. Physically, 
$J(t)$ is the  energy flux toward 
the external reservoirs  integrated over a time $t_s$.

In order to fix the value of the threshold, we computed the average number
of plateaus $\langle N_p \rangle$  as a function of $q$.
We found that there is always a range of $q$ values where $\langle N_p \rangle$ 
displays a broad maximum. Therefore, we systematically chose the value 
of $q$ in such a way as to minimize the variation of the number  
of steps. According to this prescription, our procedure identified 
approximately the same number of
relaxation stages at each realization. The average value of the
number of plateaus was found to depend primarily on the type of
potential energy of the lattice, while displaying little
sensitivity on the initial energy density. For example, with $N=160$ 
the average value of the relaxation steps before reaching
the quasi-stationary state in the NN lattices is 
$\langle N_p \rangle \approx 11 \pm 1$, while almost
twice as much relaxation stages are needed in the presence of
on-site term, $\langle N_p \rangle \approx 26 \pm 1$. We
infer that the latter system  could be characterized
by a much rougher energy landscape.

\begin{figure}[t!]
\centering
\includegraphics[width=10truecm,clip]{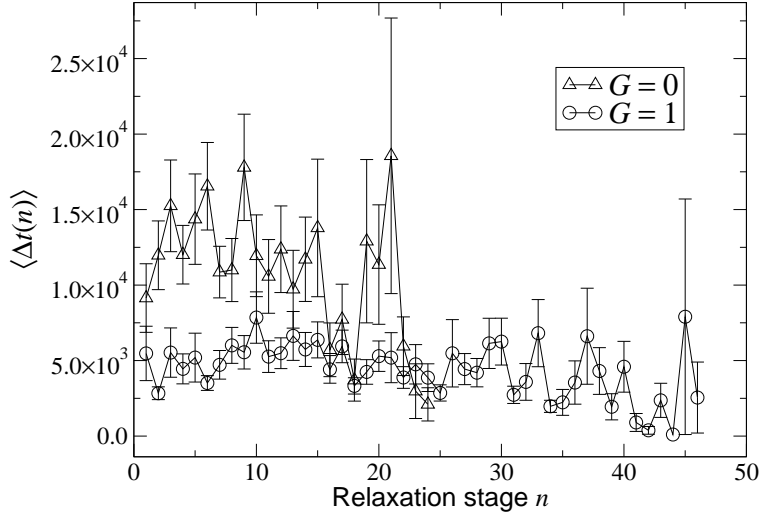}
\caption{Plot of the average duration of the plateaus $\langle
\Delta t_n \rangle$ vs order of the relaxation stage $n$ for
$G=0$ and $G=1$. Parameters are: $E(0)/N = 8$,
$N=160$, $\gamma=0.1$.} \label{dtmed}
\end{figure}

Since we do not know whether the process at hand is a stationary
one, we first checked how the averages depend on time, i.e. on the
index $n$. In Fig.~\ref{dtmed} we see that the averages $\langle
\Delta t_n \rangle$ are almost independent of the order of the
relaxation $n$ for both $G=0$ and $G \neq 0$. At least to
a first approximation, it is thus legitimate to assume that the
process is a stationary one. It is important to stress that the
above observation rules out the conjecture that the slow decay
kinetics of the system may be the result of a progressive increase
of the residence times in metastable states visited during relaxation.

\begin{figure}[t!]
\centering
\subfigure[]{
\includegraphics[width=10. truecm,clip]{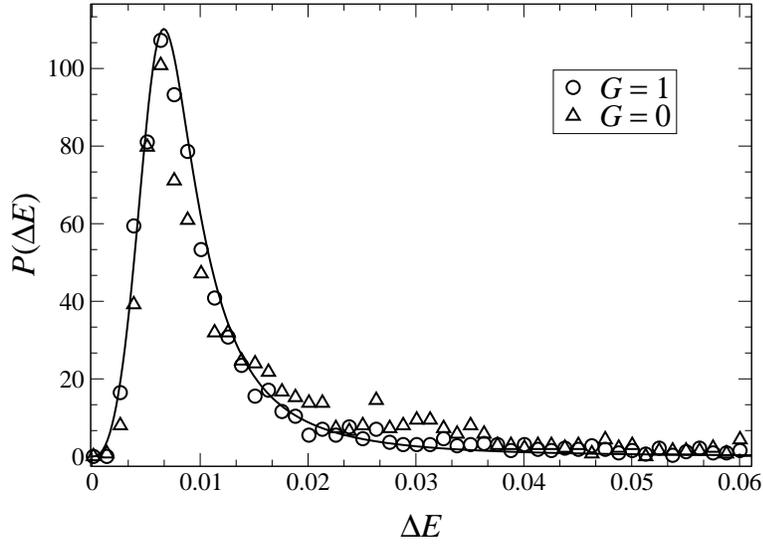}}
\subfigure[]{
\includegraphics[width=10. truecm,clip]{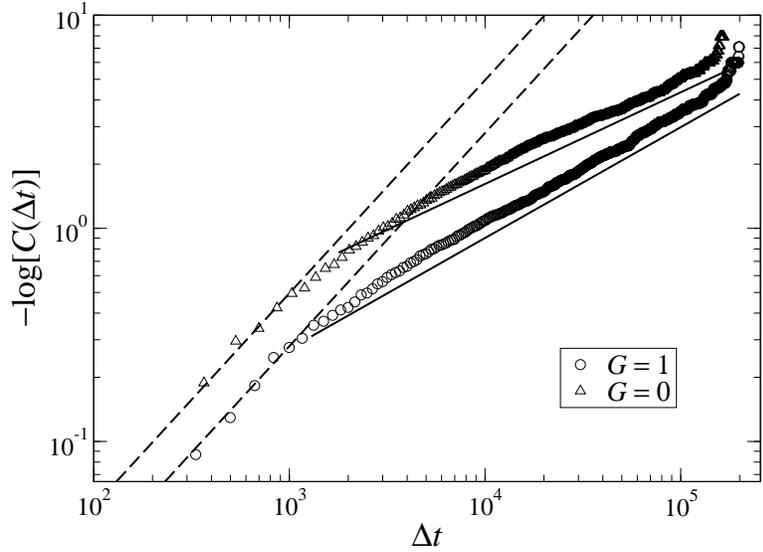}}
\caption{Statistics of the relaxation process. 
(a) Normalized histograms of energy jumps $\Delta E$ for 
$G=0$ (triangles) and $G=1$ (circles) and two--parameter fit
with the expression~(\ref{PDe}).
(b) Cumulative distribution of the plateau durations for 
$G=0$ and $G=1$ (symbols). The dashed lines are pure exponential fits
of the portion $\Delta t<\Delta t^\ast$. The solid lines
are stretched exponential fits in the region  $\Delta t > \Delta t^\ast$
(slightly shifted for the sake of clarity). The best --fit values of
the exponent $\beta$ are: $\beta=0.43$ ($G=0)$ and $\beta=0.52$ ($G=1)$. 
The series $\{ \Delta t \}$, $\{\Delta E\}$  have
been collected on a total of 99 initial conditions. 
Parameters are: $E(0)/N = 8$, $N=160$, threshold  $q=0.001$.}
\label{f:De-stat}
\end{figure}

The next step is to investigate the  probability distributions of
the lifetimes $\mathcal{P}(\Delta t)$ and energy jumps 
$\mathcal{P}(\Delta E)$. 
In Fig.~\ref{f:De-stat} (a) we show the histograms of the energy 
jumps for the two cases $G=0$ and $G \neq 0$.
We found that the data can be fitted by the empirical 
normalized function
\begin{equation}
\label{PDe}
\mathcal{P}(\Delta E) = \frac{\alpha}{\Delta E}
                        \left(
                          \frac{\Delta E_0}{\Delta E}
                        \right)^{\alpha}
                        e^{-(\Delta E_0/\Delta E)^{\alpha}}
                        \quad .
\end{equation}
Hence, the distribution of $\mathcal{P}(\Delta E)$ is
exponentially small for small values of $\Delta E$ and decays
following a power law  of the kind $1/\Delta E^{\alpha+1}$ for
large values of the energy jumps. In the case shown in 
Fig.~\ref{f:De-stat}, we found $\alpha
\approx 1.6$, $\Delta E_0 \approx 0.0081$ ($G=0$), and
$\alpha \approx 1.8$, $\Delta E_0 \approx 0.0074$ ($G = 1$).

We turn now to the distribution of plateaus $\mathcal{P}(\Delta t)$.
In practice, it is a difficult task  
to construct the histograms from the measured time series,
since the latter span a wide range (3 decades) yielding exceedingly 
large fluctuations.   A way around this problem is to 
construct the cumulative (integrated) distributions $\mathcal{C}(\Delta t)$, 
that can be obtained from the numerical series by inverting the axes
in a simple rank--size plot of the data.
Remarkably, we found that the statistics of the time intervals follows an inherently slow
stretched exponential distribution law. This finding is
illustrated  in Fig.~\ref{f:De-stat} (b) for both the $G=0$ and $G \neq 0$
cases, where we plot the quantity $-\log [\mathcal{C}(\Delta t)]$. 
More precisely, we see that the first part of the distribution is
a pure exponential (straight lines with slope one in
Fig.~\ref{f:De-stat} (b)), while the tails follow a stretched exponential
law. This means that the  probability
distribution $\mathcal{P}(\Delta t)$ (the derivative of 
$\mathcal{C}(\Delta t)$) is
\begin{equation}
\label{PDt}
\mathcal{P}(\Delta t) 
            \propto
            \begin{cases}
              \exp[-\Delta t/\tau_1] & \Delta t \ll \Delta t^* \\
              \exp[-(\Delta t/\tau_2)^\beta] & \Delta t \gg \Delta t^*
            \end{cases} 
\end{equation}
up to power--law corrections. 
The stretched exponential portion of the curves sets in at times
longer than $\Delta t^* \approx 10^3$, which is systematically
below the average, approximately constant duration of each
relaxation stage. This observation suggests that the statistics of
the plateau duration is dominated by the stretched exponential
tail. Such intrinsically fat--tailed distribution of a series
of relaxation events is thus directly related to the slow energy 
decay.

In general, the origin of a stretched exponential relaxation
is traced back to a superposition of many different time scales,
distributed according to given weights~\cite{Aubry2,Palmer}.
In our context, however, it is not clear how to identify
such a hierarchical set of degrees of freedom, that would be
responsible for the observed slow kinetics. It is thus 
legitimate to explore the possibility of alternative explanations.
An interesting perspective has been presented by
Bunde {\em et al.} in a recent paper~\cite{Bunde}, in connection 
with the statistics of extreme events.
According to the authors, a stretched exponential law may 
rule the distribution of intervals between consecutive 
rare events provided the original time series displays 
power--law correlation. More precisely, they consider a stochastic
time series $x(t)$, whose autocorrelation decays by construction
as $t^{-\gamma}$ with $0<\gamma<1$. In this context, rare events
are defined by the condition $x(t)>q$, where $q$ is a pre--assigned 
threshold. Their numerical analysis shows
that the distribution of the time intervals $\Delta t$ separating extreme
events decays as $\exp[-\Delta t^{\gamma^{\, \prime}}]$, with 
$\gamma^{\, \prime} \approx \gamma$.

\begin{figure}[t!]
\centering
\includegraphics[width=10truecm,clip]{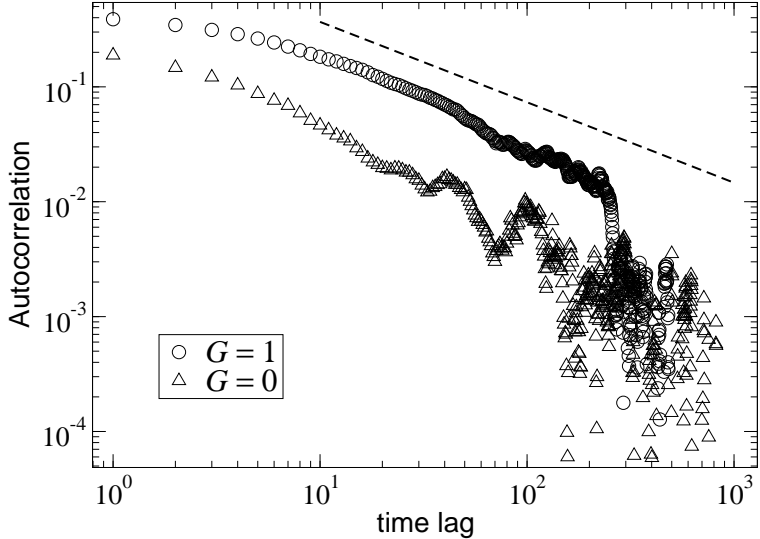}
\caption{\label{f:flux} Autocorrelation of the time series $J(t)$ averaged
over 99 initial conditions for 
$N=160$, $\gamma=0.1$, $E(0)/N=8$. The dashed line is an inverse 
power law  with exponent 0.7.}
\end{figure}

It appears natural to extend this kind of description to the process
of energy  release in our system: each energy step $\Delta E$, marking the 
collapse of a rotobreather,  can be regarded as a ``rare event '' 
(sharp peak)  in the series of energy flux $J(t)$ (see Fig.~\ref{f:enall}). 
In order to validate this analogy, we checked that long--range 
correlations do exist in the series $J(t)$ (Fig.~\ref{f:flux}).
Within the statistical fluctuations, the autocorrelation function
of $J(t)$ decreases approximately as an inverse power law, whose
exponent is smaller than one ($\approx 0.7$). 
This value is of the same order  of the 
exponents reported in Fig.~\ref{f:De-stat} (b).
As a matter of fact, the statistical fluctuations prevent us from obtaining 
a better estimate. Furthermore, the distribution of plateau durations
are only asymptotically stretched--exponential, thus introducing a further
source of error. Therefore, to make our analogy more quantitative
further  numerical work should be carried out.

We note that a similar  analysis of the correlations in the
series of the energy flux in the Fermi--Pasta--Ulam chain does not
yield a power--law decay. In fact, the autocorrelation function of $J(t)$
vanishes exponentially.  This is consistent with the absence of stretched 
exponential relaxation in this model~\cite{noi}, thus suggesting that the 
microscopic mechanisms operating in the  chain of coupled rotators are 
substantially different.

In order to provide further support to our interpretation, we have set
up a  Monte Carlo procedure which generates stochastic series
of $N_p$ energy jumps and plateaus as in Eq.~(\ref{piecewdec})
according to given distributions 
$\mathcal{P}(\Delta E)$ and $\mathcal{P}(\Delta t)$.
We have then computed the average over such a set of realizations
according  to Eq.~(\ref{piecewdec}), where $N_p$ can be fixed close to
its empirical average value.
We found that, by using  a distribution of the form~(\ref{PDe}) and a
stretched exponential law for the $\mathcal{P}(\Delta t)$,
we could recover a
stretched exponential decay of the average energy by just
tuning at most two of the free fitting parameters
$\Delta E_0,\alpha,\tau_2,\beta$. 
For example, by fixing $N_p  = 15$, $\Delta E_0 = 0.0081$ and 
$\alpha=1.7$ in the case $G=0$, $E(0)/N=8$, the best--fit value of the exponent 
$\sigma$ from the Monte Carlo simulation ($\sigma = 0.42$) is found to be 
in very good agreement with the one directly fitted on the simulation data
(see Table~\ref{t:exp}).

\section{The effect of isotopic disorder\label{sec:4}}
%
%
In this section we investigate the role of disorder on the relaxation dynamics
of the sine--lattice system. More specifically, we consider model~(\ref{onsite})
with randomly distributed momenta of inertia, that is either $I_i=R$ 
(with probability $\rho$) or $I_i=1$ (with probability $1-\rho$).
This choice corresponds to a 4--methyl--pyridine quasi--1D chain with random insertions
of deuterated methyl groups, that has been studied experimentally by Fillaux and 
co--workers~\cite{FC,FC3}. Hence, we fix $R = 2$, that corresponds to the ratio of momenta of inertia reported 
in Ref.~\cite{FC}.

  The motivation of this analysis is twofold. First, the joint effect of non--linearity
and disorder on the non--equilibrium properties of discrete lattices are far from
being satisfactorily understood~\cite{KA}. Second, the disordered sine--lattice model has 
been used to interpret data from inelastic neutron scattering experiments~\cite{FC}.
More specifically, it has been shown that the effects of isotopic disorder 
on the dynamics rapidly disappear upon increasing the temperature.
This observation has been explained quantum--mechanically as a
transition to the classical regime, thus suggesting that the effects of disorder
should be negligible in the latter case.
By the same token, one may expect that the relaxation dynamics in the classical
system should display
weak sensitivity on the concentration of impurities.

\begin{figure}[t!]
\centering\includegraphics[width=10cm, clip]{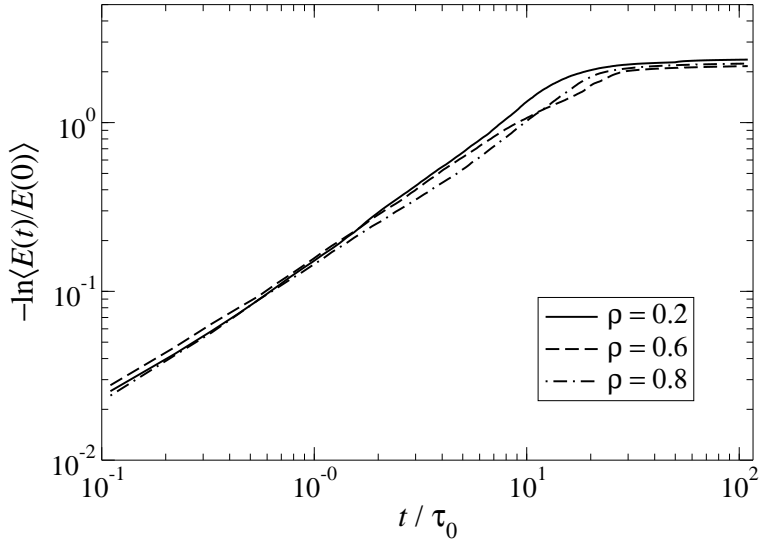}
\caption{Decay of the normalized total energy  in a chain with isotopic 
disorder.  We plot $-\ln \langle E(t)/E(0) \rangle$ in log--log scale
with $E(0)/N = 4.6$ and for three different 
values of the concentration $\rho$. 
The data have been averaged over a set of  30 initial conditions.
Parameters are $N=100$, $G=1$, $\gamma=0.1$.}
\label{f:disorder}
\end{figure}

  Our simulations confirm the above hypothesis. In Fig.~\ref{f:disorder} we show
that the energy decay is hardly affected by the value of $\rho$. Remarkably, 
the curves almost overlap and display both a wide region of stretched exponential 
behavior and  the saturation to a residual state.
Accordingly, the estimated exponents are fairly insensitive to the concentration 
of impurities.
Moreover, we checked that the energy which is left in the residual state is approximately
independent of $\rho$, with deviations of at most 10 \% from the average.

\section{Conclusions\label{sec:5}}

The study of the energy relaxation in models of classical rotators has allowed
us to identify a possible mechanism yielding the stretched exponential
behavior, observed also in other one--dimensional models
\cite{Aubry1,Aubry2,KA}. The static nature of localized breather states, which
are spontaneously formed upon cooling the lattice from its boundaries, is one
of the basic ingredients. Indeed, static localized solutions can form
``dynamical barriers" that are almost decoupled from their neighbours and thus
segregate an internal hot lattice region from the cooled boundaries. The energy
initially fed into the system remains trapped there, until the rotobreathers
are abruptly destroyed.  It may be argued that this sudden death must originate
from a sufficiently large resonant fluctuation emerging from the hot core. In our
approach, we have employed a statistical description, which does not make
any explicit reference to specific dynamical mechanisms. 

Accordingly, we showed that the destruction of
boundary rotobreathers may be regarded as a ``rare event''. The
resulting
statistics depends on two main factors: the energy of the rotobreathers and the
probability that in a given time interval the hot region may produce an
excitation capable of annihilating them. In our numerical analysis
we can just observe the effects of this complicated interaction, which results 
in a stretched--exponential law for the energy decay.

It may be argued that the mechanism of energy trapping must
be rather peculiar to the 1D nature of the model. Indeed, we
expect that in higher-dimensional lattices this phenomenon
may not occur at all, and that the characteristic stepwise decay
of energy could be replaced by a smoother decrease.
Some preliminary simulations of a 2D arrays of rotators confirm this
idea. In particular, we observe that the residual state resembles the
one of the 2D Fermi--Pasta--Ulam model - a collection of static 
breathers randomly arranged on the lattice \cite{chaos}.

Altogether, we have seen that the possibility of easily exciting long--lived 
rotobreathers is responsible for the slow kinetics. It should be recalled that these
very same objects have been invoked to explain the normal conduction in the NN 
model~\cite{therm}. In this sense, our results complement those observations 
and confirm the  essential role of nonlinear excitations in the 
out--of--equilibrium
properties of many--body systems.

%

\section*{Acknowledgements}

This work has been partially supported by the INFM--PAIS project {\it
Transport phenomena in low--dimensional structures} and by the EU
network LOCNET, Contract No. HPRN-CT-1999-00163. 
F. P. acknowledges the hospitality during his stay at the 
``Centro interdipartimentale per 
lo Studio delle Dinamiche Complesse'' of the University of Florence
(CSDC, \texttt{www.complex.unifi.it}).

\end{document}